\documentclass[12pt,toc]{JHEP}
\usepackage{graphicx} 

\def\be{\begin{equation}}
\def\ee{\end{equation}}
\def\ba{\begin{array}{l}}
\def\ea{\end{array}}
\def\bea{\begin{eqnarray}}
\def\eea{\end{eqnarray}}

\def\fig#1{Fig \ref{#1}} 

\def\del{\partial}

\def\p{p^{\bar x}}

\def\K{{\mathbf K}}

\def\nn{\nonumber\\}

\def\K3{{\bf K3}}
\def\journal#1&#2(#3){\unskip, \sl #1\ \bf #2 \rm(19#3) }
\def\andjournal#1&#2(#3){\sl #1~\bf #2 \rm (19#3) }

\def\hat{\widehat}

\def\tilde{\widetilde}

\def\frac#1#2{{#1\over#2}}

\def\inbar{\,\vrule height1.5ex width.4pt depth0pt}
\def\IC{\relax\hbox{$\inbar\kern-.3em{\rm C}$}}
\def\IR{\relax{\rm I\kern-.18em R}}
\def\IP{\relax{\rm I\kern-.18em P}}

\catcode`\@=11
\def\slash#1{\mathord{\mathpalette\c@ncel{#1}}}
\def\CC{{\cal C}}

\def\LL{{\cal L}}

\def\underrel#1\over#2{\mathrel{\mathop{\kern\z@#1}\limits_{#2}}}

\catcode`\@=12

\def\det{{\rm det}}

\def \sinh{{\rm sinh}}
\def \cosh{{\rm cosh}}

\def\det{{\rm det}}
\def\exp{{\rm exp}}

\def \ov {\over}
\def \p {\partial}
\def \ha {{1 \ov 2}}

\def \om {\omega}

\def \ep {\epsilon}

\def\le{\left}
\def\ri{\right}

\def\IL{\relax{\rm I\kern-.18em L}}
\def\IH{\relax{\rm I\kern-.18em H}}
\def\IR{\relax{\rm I\kern-.18em R}}
\def\IC{\relax\hbox{$\inbar\kern-.3em{\rm C}$}}

\def\sector#1#2{\ {\scriptstyle #1}\hskip 1mm
\mathop{\opensquare}\limits_{\lower
1mm\hbox{$\scriptstyle#2$}}\hskip 1mm}

\def\tsector#1#2{\ {\scriptstyle #1}\hskip 1mm
\mathop{\opensquare}\limits_{\lower
1mm\hbox{$\scriptstyle#2$}}^\sim\hskip 1mm}

\def\nn{\nonumber}

\title{R-charged $AdS_{5}$ black holes and large N unitary matrix models}

\author{  
Pallab Basu \\ 
\email{pallab@theory.tifr.res.in} 
}
\author{
Spenta R. Wadia \\ 
\email{wadia@theory.tifr.res.in} \\
}
\author{ Tata Institute of Fundamental Research, \\
Homi Bhaba Road, Mumbai-400005, India \\
}

\preprint{TIFR/TH/05-24}

\abstract{
Using the AdS/CFT, we establish a correspondence between the intricate thermal phases of R-charged $AdS_{5}$ blackholes and the R-charge sector of the N=4 gauge theory, in the large N limit. Integrating out all fields in the gauge theory except the thermal Polyakov line, leads to an effective unitary matrix model. In the canonical ensemble, a logarithmic term is generated in the non-zero charge sector of the matrix model. This term is important to discuss various supergravity properties like $i)$ the non-existence of thermal $AdS$ as a solution, $ii)$ the existence of a point of cusp catastrophe in the phase diagram and $iii)$ the matching of saddle points and the critical exponents of supergravity and those of the effective matrix model.
}

\begin{document}

\section{Introduction}
\par
 
The AdS/CFT correspondence\cite{Maldacena:1997re} implies that the phases of string theory can be studied by studying those of the dual gauge theory(\cite{Gubser:1998bc,Witten:1998zw,Witten:1998qj}). In the case of type $IIB$ string theory in $AdS^{5}\times S^{5}$ the phases can be studied using the large N limit of a unitary matrix model(\cite{Sundborg:1999ue,Polyakov:2001af,Aharony:2003sx,Liu:2004vy})\footnote{Phases of large N gauge theory is also discussed in \cite{Hallin:1998km,Schnitzer:2004qt}}. The unitary matrix is the finite temperature Polyakov loop which does not depend on the points of $S^3$. This fortunate circumstance is due to the fact that in the Hamiltonian formulation, N=4 SYM theory at a given time slice, is defined on the compact space $S^3$ and the $SO(6)$ scalars are massive because of their coupling to the curvature of $S^3$. These facts imply that, in principal, one can integrate out almost all the fields and obtain an effective theory of the zero mode of the gauge potential $A_0$. 

Using this method a detailed correspondence of the critical points of the gauge theory effective lagrangian and the critical points of supergravity (discussed by Hawking and Page\cite{Hawking:1982dh}) can be constructed at the leading order of the $1/N$ expansion. These are $AdS_5$ and the small and big black holes (as \cite{Alvarez-Gaume:2005fv}  we refer to these as SSB and BBH.) It turns out that in the gauge theory these critical points are in the gaped phase, where the density of eigenvalues vanishes in a finite arc of the circle around which the eigenvalues are distributed. The closing of the gap, corresponds to the Gross-Witten(GW) phase transition(\cite{Gross:1980he,Wadia:1979vk,Wadia:1980cp}). In a window around this transition, the supergravity description of string theory is likely to smoothly cross over into a description in terms of heavy string modes. \footnote{In \cite{Alvarez-Gaume:2005fv}, besides the vicinity of the Gross-Witten transition, authors also studied $1/N$ corrections and presented formulas for the partition function in the vicinity of blackhole nucleation and the Hagedorn transition.}

In this paper we extend the discussion of the correspondence between R-charged $AdS_{5}$ blackholes(\cite{Cvet1,Cvet2,Chamblin:1999tk,Chamblin:1999hg,Hawking:1999dp}), and the effective unitary matrix model(\cite{Gross:1980he,Wadia:1979vk,Wadia:1980cp,Halpern:1982rd}). R-charged black holes are known to have a rich phase structure in the canonical and grand canonical ensemble. In the canonical ensemble the fixed charge constraint, contributes an additional logarithmic term $\log(TrUTrU^{\dagger})$  involving the order parameter, to the gauge theory effective action. This term is crucial for matching with supergravity. We analyze the implications of this term in the large N limit and compare with the various supergravity properties like the existence of only blackhole solutions in the canonical ensemble and also the existence of a point of cusp-catastrophe in the phase diagram.
 
\par 
The plan of this paper is as follows. In section \ref{Supergravity} we give a brief review of charged $AdS_5$ blackholes. In section \ref{sec:free} and section \ref{sec:small} we discuss the effective action of the gauge theory at zero and small coupling, in the fixed charged sector. At zero coupling there is exactly one saddle point and the value of $TrUTrU^{-1}$ at the saddle point is always non-zero. For a small positive coupling there are two stable and one unstable saddle points, all with a non-zero value of $TrUTrU^{-1}$. They merge at the $GW$ point. In section (\ref{sec:simmod}) we discuss the model effective action at strong coupling. Here too, there are three saddle points,two stable(I,III) and one unstable(II). In the region $\rho>\frac{1}{2}$, $I$ and $III$ can be identified with a stable small blackhole and stable big blackhole respectively. Saddle point $II$ is identified with the small unstable black hole. The merging of saddle points leads to critical phenomenon whose exponents can be calculated and shown to agree with supergravity. This is discussed in section \ref{Unin}. We have also calculated the $o(1)$ part of the partition function near the critical point. 

\section{R-charged blackholes in $AdS_{5}$ and critical phenomena }\label{Supergravity}
The R-charged $AdS_{5}$ black hole and relevant phase structure were discussed by A.~Chamblin, R.~Emparan, C.~V.~Johnson and R.~C.~Myers (\cite{Chamblin:1999hg},\cite{Chamblin:1999tk}). Here we review their result. The Einstein--Maxwell--anti--deSitter (EMadS$_{n+1}$) action may be written as
\begin{equation}
I = -\frac{1}{16{\pi}G} \int_{M} d^{n+1}x \sqrt{-g} \left[\tilde R - F^2 +
\frac{n(n-1)}{R^2}\right]\ ,
\label{actionjackson}
\end{equation}
where $\tilde R$ is the Ricci scalar and $R$ is the characteristic length scale of $AdS$.  
The metric of the Reissner--Nordstr\"om--anti--deSitter (RNadS)
solution is given in static coordinates by

\begin{equation}
ds^2 = -(1 - \frac{m}{r^{n-2}} + \frac{q^2}{r^{2n-4}} + \frac{r^2}{R^2}\ )dt^2 + \frac{dr^2}{1 - \frac{m}{r^{n-2}} + \frac{q^2}{r^{2n-4}} + \frac{r^2}{R^2}\ } +r^{2}d{\Omega}^2_{n-1}\ ,
\label{BlackHole}
\end{equation}
\noindent
The parameter
$q$ is proportional to the charge
\begin{equation}
Q=\sqrt{2(n-1)(n-2)}\left({\omega_{n-1}\over 8\pi G}\right)q\ 
\label{thecharge}
\end{equation}
 and $m$ is proportional to the ADM mass $M$ of the blackhole.
\begin{equation}
M={(n-1)\omega_{n-1}\over 16\pi G}m\ 
\end{equation}
$\omega_{n-1}$ is the volume of the unit $(n{-}1)$--sphere, and the gauge potential is given by
\begin{equation}
A_{0}=\left(-{1\over {\sqrt{2(n-2)\over n-1} }}{q\over r^{n-2}}+\Phi\right)
\label{pure}
\end{equation}
where $\Phi$ is the electrostatic potential difference between the black hole horizon and infinity.
\par
For n=4 the solution (\ref{BlackHole}) can be considered as a rotating black hole in $AdS_{5}\times S^{5}$ with angular momentum in the internal space $S^{5}$.\footnote{The EMadS system described here may be thought as the dimensional reduction of $AdS_{5}\times S^{5}$ to $AdS_5$. Generally introducing angular momentum in $S^{5}$ will distort $S^5$. This distortion is not taken into in the EMadS reduction \cite{Behr,Kraus:1998hv}. We thank S.Trivedi for pointing this out to us. However these details do not change our main consideration.} The symmetry group of $S^5$ is $SO(6)$ and the black hole we are discussing has equal $U(1)$ charges for all the three commuting $U(1)$ subgroups of $SO(6)$, the  R-symmetry group of the $N=4$ $SYM$. Hence we are dealing with a system which has the same chemical potential $\mu$ for all three $U(1)$ charges in the grand canonical ensemble or equivalently three fixed equal U(1) charges in the canonical ensemble.

\subsection{Equation of state}\label{sec:state}

In order to discuss the thermodynamics, we consider the Euclidean continuation ($t{\to}i\tau$) of the solution, and identify the imaginary time period $\beta$ with the inverse temperature. Using the formula for the period, $\beta{=}{4\pi}/{V^{\prime}(r_{+})}$ (for a review see \cite{David:2002wn}), we get

\begin{equation}
\label{betaform}
\beta = \frac{4{\pi}l^{2}r_{+}^{2n-3}}{nr_{+}^{2n-2} +
(n-2)l^{2}r_{+}^{2n-4} - (n-2)q^{2}l^{2}}\ .
\end{equation}

\noindent This may be rewritten in terms of the potential as:
\begin{equation}
\beta = \frac{4{\pi}l^{2}r_+}{(n-2)l^{2}(1-c^2\Phi^2)+nr_+^2}\ .
\label{betaformtwo}
\end{equation}

The condition for euclidean regularity used to derive (\ref{betaform}) is equivalent to the condition that the black hole is in thermodynamical equilibrium. The equation (\ref{betaform}) may therefore be written as an equation of state $T{=}T(\Phi,Q)$. From this equation of state we see that for fixed $\Phi$ we get two branches, one for each sign, when the discriminant under the square root is positive\cite{Chamblin:1999hg}. For fixed $Q$, $T(\Phi)$ has three branches for $Q{<}Q_{crit}$ (Let us call them $I$, $II$, $III$) and one for $Q{>}Q_{crit}$. The critical charge is determined at the "point of inflection'' by $\left(\partial Q/\partial\Phi\right)_T{=}\left(\partial^2 Q/\partial\Phi^2\right)_T{=}0.$

\par
\begin{figure}[h!]
\begin{center}
\includegraphics[height=2in]{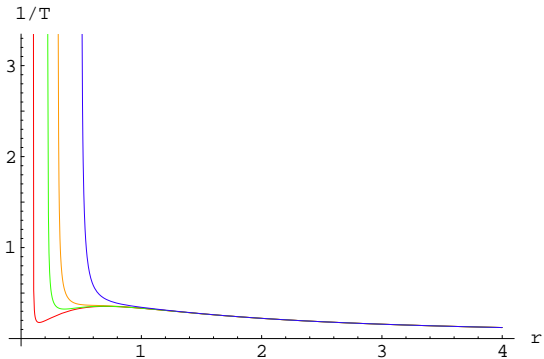}    
\end{center}
\caption{Plot of $\beta(r_{+})$ for q increasing from the left. The third graph from the left is for $q_{crit}$ }
\label{fig:beta}
\end{figure}

 The qualitative features of $\beta(r_+)$ for varying $q$ are shown in
Fig \ref{fig:beta}.
There is a critical charge, $q_{\rm crit}$, below which there are three solutions for $r_+$ for a range of values of $T$,
corresponding to small ($I$), unstable($II$), and large ($III$) black holes. For fixed $q<q_{crit}$ only branch $I$ is available at low temperatures. At $T=T_{02}(q)$, there is a nucleation of two new solutions, $II$ the unstable small black hole solution, and $III$ the stable big black hole solution. As the temperature is increased further, the black hole $II$ approaches black hole $I$ and at $T=T_{02}(q)$ the two solutions merge. 
\par
  As $q$ is increased further, $T_{02}$ increases, whereas $T_{01}$ decreases. At $q=q_{crit}$ we have $T_{crit}=T_{01}=T_{02}$. At $q_{crit}$ and $T_{crit}$ all three solutions merge. In the language of catastrophe theory this is a cusp catastrophe. As we increase $q$ beyond $q_{crit}$ there will be just one solution for all temperatures $T$. 
\par
 We wish to take note of some properties of the phase diagram.

\begin{enumerate}
\item Thermal $AdS$ is not a solution and all three branches of the solution represent black holes.
\item There exists a critical point where the three solutions of the system merge. It is a point of fold catastrophe.   
\end{enumerate}

\subsection{Critical Phenomena}
\par 
 The critical point ($Q_{cit}$,$T_{crit}$) may be approached from various directions in the parameter space. If we set $T=T_{crit}$, then the equation determining $(r-r_{crit})$ takes the form ($r_{crit}$ is the radius of the critical black hole)
\bea
(r-r_{crit})^{3}= \CC (Q-Q_{crit})
\eea
$\CC$ is a numerical constant. The critical exponent here is $\frac{1}{3}$, since 
\be
(r-r_{crit}) \propto (Q-Q_{crit})^{\frac{1}{3}}
\ee
\par 
 As discussed in \cite{Chamblin:1999hg} (Fig16), the critical point may also be approached through the coexistence line in the parameter space. The coexistence line is the line with the property 
\be
\nn S_{I}=S_{III}
\ee     
for the parametric range $q<q_{crit}$. It is the line where the Hawking-Page (first order) transition from the small black hole to the big black hole takes place. As we approach the critical point through this line, we have the relation 
\bea
(r_{I}-r_{II}) \propto (T-T_{crit})^{\frac{1}{2}}
\eea 
 \par
In the following we will present an understanding of these properties. Before we do the matching with supergravity we would like to present a discussion of the gauge theory in the limit when $\lambda=g^2 N=0$ and also when $\lambda << 1$. In these cases we of course can not compare with supergravity which requires $\lambda >> 1$. 
 
\section{Free YM theory}\label{sec:free}
\subsection{Effective action with chemical potential}
In this section we  briefly review (see \cite{Liu:2004vy},\cite{Aharony:2003sx}) the effective action for a
free $SU(N)$ Yang-Mills theory (with adjoint matter) on a compact manifold $\Im$ in the large $N$ limit. The basic idea is to integrate out all fields in the theory except for the zero mode of the Polyakov line. The partition function is then reduced to a single unitary matrix integral.

Expanding all fields in the gauge theory in terms of harmonics on $\Im$, the theory reduces to a zero dimensional problem of free $N \times N$ Hermitian matrices  
 \bea
\LL = \ha \sum_a Tr \le[ (D_t M_n)^2  - \om_n^2 M_n^2 \ri]
\label{simlad}     
\eea
The sum in (\ref{simlad}) )is over all field types and their Kaluza-Klein modes on $\Im$.
$\om_n$ is the frequency of each mode. The covariant derivative in (\ref{simlad}) is 
\be
D_t M_n = \p_t M_n - i [A_{0}, M_n]
\ee
$A_{0}$ comes from the zero mode (i.e. the mode independent of
coordinates on $\Im$) of the time component of the gauge field and
is not dynamical. The partition function of the theory at finite temperature
can be written as a unitary matrix integral by integrating out all fields in (\ref{simlad}) except for $A_{0}$. Hence we have
 \bea
 Z & = \int DU \, \prod_n
 \le(\det_{adj} \le(1- \ep_n e^{-\beta \om_n} U \ri)\ri)^{-\ep_n} 
 \eea
where $U=\exp(i\beta A_{0})$ is a $U(N)$. $\det_{adj}$ denotes the
determinant in the adjoint representation and $\ep_n =  1$ ($-1$)
for bosonic (fermionic) $M_n$. The above equation can be 
 expressed as
 \bea
 Z =  \int DU \, \exp (\sum_{n=1}^\infty {z_{n}(\beta) \ov n})
 Tr U^n Tr U^{-n} )
\eea
where
\bea
 z_n(\beta) = z_B (n \beta) +(-1)^{n+1} z_F (n\beta) .
 \eea
Here $z_B(\beta), z_F (\beta) $ are the single particle partition
functions of the bosonic and fermionic sectors respectively ( see \cite{Aharony:2003sx} for the explicit formulas for $z(\beta)$ in various gauge theories).
 
\par
If we introduce a chemical potential $\mu=(\log m)$, the formula for the partition function changes to
\be
Z[\beta,\mu]=\int DU e^{(\sum_{n=1}^\infty {z_{n}(\beta,\mu) \over n} TrU^{n}TrU^{-n})} 
\ee
where
\be
z_n(\beta,\mu)= z_B (n \beta,n \mu) +(-1)^{n+1} z_F (n\beta ,n \mu)
\ee
$z_B(\beta,\mu), z_F (\beta,\mu) $ are now the single particle partition functions with chemical potential $\mu$. Hence 
\bea
z_B(\beta,\mu) &=& \sum_{bosons}\exp(-E_{i}\beta+Q_{i}\mu) \\
z_F(\beta,\mu) &=& \sum_{fermions}\exp(-E_{i}\beta+Q_{i}\mu) 
\eea

$Q_{i}$ is the charge of the state whose energy is $E_{i}$ (see \cite{Aharony:2003sx}).

\par
As we have already mentioned, we wish to describe a system which has the same chemical potential $\mu$ for all three $U(1)$ charges of the R-symmetry group $SO(6)$ in grand canonical ensemble. Equivalently, we can work with fixed and equal values of the $U(1)$ charges in the canonical ensemble.   
\par
Let us for simplicity confine ourselves to the bosonic sector of the $N=4$ SYM theory. The gauge fields have no R-charge. The six scalars $\phi_{i}$ $(i=1,...,6)$ can be grouped in pairs of two. We define 
\bea
\phi^{+}=\phi^{1}+i\phi^{2} ,
&& 
\phi^{-}=\phi^{1}-i\phi^{2}
\eea
(We can similarly define complex fields for the other two pairs.) $\phi^{\pm}$ have charge $ \pm 1$ for each of the three commuting $U(1)s$ of $SO(6)$. Hence, if we consider the single particle partition function for these fields, it will be 
\be
z[x,\mu]=(exp(+\mu)+exp(-\mu)) z[x,0]/2
=\cosh(\mu)z[x,0]
\label{chpt}
\ee 
where $z[x,0]$ is the single particle partition function without any chemical potential. 

\subsection{Canonical Ensemble}
      We will now discuss the free gauge theory partition function for a  canonical ensemble with constant charge $Q_{0}$, by introducing a delta function $\delta(\hat Q-Q_{0})$ in the functional integration of the gauge theory. $\hat Q=Q[\phi]$ is the corresponding functional for the charge which we want to keep fixed. In our case $\hat Q$ is just the functional for R-charge in gauge theory. 
\par
The fixed charge partition function is defined by
 
\bea  
  Z(\beta,Q_{0}) &=& \int DX e^{\int_{0}^{\beta} S[X]} \delta(\hat Q-Q_{0})  \\
 &=& \int DX e^{\int_{0}^{\beta} S[X]} \int \exp(i\mu \hat Q)\exp(-i\mu Q_{0})d\mu  \\
&=& \int d\mu \exp(-i\mu Q_{0})(\int DX e^{\int_{0}^{\beta} S[X]}e^{i\mu \hat Q} ) \\
&=& \int d\mu \exp(-i\mu Q_{0})\int DU \exp(\Sigma z_{n}[\beta,i\mu]TrU^{n}TrU^{-n}) 
\label{chargefix1}
\eea 
where $z_{n}[\beta,i\mu] = z^{V}_{n}[\beta,0]+cos(\mu)z^{S}_{n}[\beta,0]+\cos(\frac{\mu}{2})z^{F}_n[\beta,0]$. 
\par
We can now make the approximation \footnote{This approximation can be thought of as a low temperature approximation. This is because at low temperatures $z^{S}_{n}[\beta,0]$ approaches zero as $e^{-\beta n}$. Hence the higher $z_{n}$ are suppressed relative to $z_{1}$. It is also true that for all temperatures,  $z_{n} < z_{1}$  and for very high temperatures we have $z_{n} \sim \frac{z_{1}}{n}$. As an example, the total contribution for all other $z_{n}$, even near hagedorn transition in free N=4 SYM  theory, is only about $7\%$ of $z_1$\cite{Aharony:2003sx}.
\par
Unitary matrix models involving $TrU^{n},n>1$ has been discussed in \cite{Jurkiewicz:1982iz,Mandal:1989ry}.
 } that $|z_{n}[x,i\mu]|$ for $n>1$ is negligible in comparison to $|z_{1}[x,i\mu]|$. Neglecting the contribution from the $n>1$ modes we arrive at a model which contains only $TrUTrU^{-1}$. Using the specific formula for $z[x,\mu]$, of the bosonic \footnote{Effect of the fermions is discussed in appendix \ref{app:fermion}.} sector ,
\bea
\nn  Z(\beta,Q_{0}) &=& \int d\mu \exp(-i\mu Q_{0})\int DU \exp( (a+c\cos(\mu))TrUTrU^{-1}) \\
\nn &=& \int DU \exp(aTrUTrU^{-1})\int d\mu \exp(-i\mu Q_{0})\exp(c \cos(\mu)TrUTrU^{-1}) \\
&=& \int DU \exp(aTrUTrU^{-1})I_{Q_{0}}(c TrUTrU^{-1}) 
\label{IntegralZ}
\eea
Here $a(\beta)=z^{V}_{n}[\beta,0]$,$c(\beta)=z^{S}_{n}[\beta,0]$ and for convenience we did not show the explicit $\beta$ dependence in the equations. $I_n(x)$ is the Bessel function.
\par
Hence we end up with a matrix model with an effective potential 
\be
S_{eff}=a(TrUTrU^{-1})+\log[I_{Q_{0}}(c TrUTrU^{-1})]
\ee
where $a>0,c>0$.
We define $\rho^{2}=TrUTrU^{-1}/N^{2}$ to get
\be
S_{eff}(\rho)=N^{2}(a\rho^{2}+(1/N^{2})\log[I_{Q_{0}}(c N^{2}\rho^{2})])
\label{seff}
\ee

It may seem that the logarithmic term is suppressed by a factor of $1/N^{2}$ and hence negligible for large N. But this need not be the case because, in the semicalssical large N limit, we must deal with a system with a charge of order $N^{2}$. Hence we define $Q_{0}=N^{2}q$ $(q \sim o(1))$. Using the asymptotic expansion of $I_{n}(nx)$ for large n, the effective action {\footnote{It should be noted that when Q=0 we should use the asymptotic expansion of $I_{0}(x)$ for large $x$. Then we get the expected answer $S_{eff}=(a+c)TrUTrU^{-1}$ which is same as a model without any constraint on charge.} becomes
\be
S_{eff}(\rho)=N^{2}(a\rho^{2}+q(\sqrt{(1+\frac{c^{2}}{q^{2}}\rho^{4})}+\log(\frac{\frac{c}{q}\rho^2}{1+\sqrt{1+\frac{c^{2}}{q^{2}}\rho^{4}}})))+O(1)
\label{Seff}
\ee
\subsection{Phase Structure}

To understand the phase diagram of this model at large N, we have to locate the saddle points of (\ref{Seff}) after including the relevant contribution from the path integral  measure depending on whether $\rho < \frac{1}{2}$ or $\rho > \frac{1}{2}$.\footnote{The term in the right hand side of equation (\ref{Tosolve2}) originates from the path integral measure over an unitary matrix.( see \cite{Goldschmidt:1979hq},Appendix of \cite{Alvarez-Gaume:2005fv}). }

Differentiating $S_{eff}(\rho)$  we get 
\bea
\nn \frac{\del}{\del\rho^{2}}S_{eff}(\rho)  &=& a+\frac{\del}{\del\rho^{2}}(\frac{1}{N^{2}}\log[I_{Q}(Q \frac{cN^{2}\rho^{2}}{Q})] \\
\nn  &=& a+b\frac{I_{Q}'(Q\frac{c\rho^{2}}{q}) } {I_{Q}(Q\frac{c\rho^{2}}{q})}  \\
 &=& a+\frac{q}{\rho^2}(1+\frac{c^{2}}{q^{2}}\rho^{4})^{\frac{1}{2}} +O(1/Q) 
\label{Tosolve1}
\eea

\par
Hence the equations to solve are 
\bea
\nn a\rho+\frac{q}{\rho}(1+\frac{c^{2}}{q^{2}}\rho^{4})^{\frac{1}{2}} &=& \rho , \rho <\frac{1}{2} \\
a\rho+\frac{q}{\rho}(1+\frac{c^{2}}{q^{2}}\rho^{4})^{\frac{1}{2}} &=& \frac{1}{4(1-\rho)},\rho>\frac{1}{2}
\label{Tosolve2}
\eea

The left side in (\ref{Tosolve2}) can be written as 

\bea
a\rho+\frac{q}{\rho}(1+\frac{c^{2}}{q^{2}}\rho^{4})^{\frac{1}{2}}=
 a\rho+c\rho+\frac{q}{\rho}\frac{1}{(1+\frac{c^{2}}{q^{2}}\rho^{4})^{\frac{1}{2}}+\frac{c}{q}\rho^{2}}
\eea

So fixing the charge gives rise to a term of type 
$\frac{q}{\rho}\frac{1}{(1+\frac{c^{2}}{q^{2}}\rho^{4})^{\frac{1}{2}}+\frac{c}{q}\rho^{2}}$ which has some important properties.
\begin{enumerate}
\item For all values of $q>0, c>0$, this term is a decreasing positive function of $\rho$ and it diverges as $\rho \rightarrow 0$.  
\item For all values of $c>0$ it is a monotonically increasing function of $q$.
\end{enumerate} 

We can now discuss the solution of this model at $N=\infty$. Let us assume that we are discussing the phase where $a(T)+c(T)<1$. This condition is valid for low temperatures since $a(T), c(T) \rightarrow 0$ as $T \rightarrow 0$. It should also be recalled that without any charge fixing the hagedorn transition occurs when $a(T)+c(T)=1$. Unlike the situation with no charge, here we have a function, on the left hand side of (\ref{Tosolve2}), which diverges as $\rho \rightarrow 0$. Hence $\rho = 0$ can not be a solution. We get only one solution at a finite value of $\rho$ which we will describe in the next paragraph. 
\par
 Equation (\ref{Tosolve2}) is solved in the region $\rho<\frac{1}{2}$ with solution 

\bea
\rho^4=\frac{q^{2}}{(1-a)^{2}-c^{2}}=\frac{q^{2}}{(1-a-c)(1-a+c)}
\label{freesol}
\eea

The self consistency condition for a solution in the region $\rho < \frac{1}{2}$ is  
\bea
\rho^4=\frac{q^{2}}{(1-a)^{2}-c^{2}}<\frac{1}{16}
\label{Selfconst}
\eea
 
 At low temperatures the condition is naturally satisfied for a small enough value of $q$. If we gradually increase the temperature (i.e. the value of $a(T)$ and $c(T)$) while keeping the value of the $q$ fixed, then the value of $\rho$ at this saddle point will increase. At some temperature $T_3(q)$, $\rho$ will become equal to $\frac{1}{2}$. Since the measure part (i.e. right hand side of (\ref{Tosolve2}) ) has a third order discontinuity at $\rho=\frac{1}{2}$, we will get a third order phase transition at the temperature $T_{3}$. From (\ref{Selfconst}) we have the following condition at $T_{3}$
\bea
\frac{q^{2}}{(1-a)^{2}-c^{2}}=\frac{1}{16}
\eea 
 If the temperature is increased beyond $T_{3}$ then we have to solve (\ref{Tosolve2}) in the region $\rho > \frac{1}{2}$. 
\par
If we increase $q$, then the value of $\rho$ at the saddle point for a fixed temperature will increase. At some $q_{3}$ we get a third order phase transition satisfying 
\bea
16q_{3}(T)^2=(a-1)^2-c^2 
\eea
Since the minimum value of $a(T)$ and $c(T)$ is zero, the maximum possible value of  $q_{3}^{2}(T)$ is $q_{crit}^{2}=\frac{1}{16}$. If we increase the $q$ beyond $q_{crit}$, the saddle point will always be confined in the parameter range $\rho > \frac{1}{2}$. Consequently as we increase the temperature from zero we will not get a third  order phase transition for  $q > q_{crit} = \frac{1}{4}$. 
       
\par
This free model, at zero gauge coupling ($l_{s}>>R$ in bulk), has some similarities with $AdS_{5}$ black holes in a fixed charge ensemble. However  unlike the three black hole branches in $AdS_{5}$, we get only one branch in the free theory. But most importantly the solution always has a nonzero value of $\rho$. 

It should be recalled that before the Hagedorn transition, a free gauge theory with zero charge has the solution $\rho=0$ \cite{Sundborg:1999ue,Aharony:2003sx}. 
\par
Some properties of the free theory will be important when we analyze the situation for the weakly coupled gauge theory. Just above the temperature $T_{3}(q)$,  the difference of the two sides of (\ref{Tosolve2}) can be expanded in the region $\rho > \frac{1}{2}$. Defining $\rho=\frac{1}{2}+x$,($x>0$) the difference is 
\bea
-\epsilon(q)x-C_{1}x^{2}
\label{expansion}
\eea
Here $\epsilon(q)>0$ and $\epsilon(q)  \rightarrow 0$ as $q \rightarrow 0$. It is important to note that $C_{1}> 2$ because the measure function (i.e. right hand side of (\ref{Tosolve2}) has a third order discontinuity at the point $\rho=\frac{1}{2}$. We will also discuss the significance of this in what follows. 


\section{Small coupling model}\label{sec:small}
 
We will now discuss the problem with a small non-zero gauge coupling $\lambda=g_{YM}^{2}N$. By AdS/CFT correspondence it corresponds to a finite string length in $AdS$. It has been shown in \cite{Alvarez-Gaume:2005fv} that by considering a phenomenological model of type 
\bea
S[TrUTrU^{-1}] = a(\lambda,T)(TrUTrU^{-1})+\frac{b(\lambda,T)}{N^2}(TrUTrU^{-1})^2 , b>0
\label{liuwa}
\eea
we can map out the possible phase diagram of type $IIB$ string theory in $AdS_{5}$.\footnote{In fact in \cite{Alvarez-Gaume:2005fv} an arbitrary convex function is considered, and shown to map out the phase diagram of $IIB$ theory in $AdS_5$. The simplified model (\ref{liuwa}) leads to similar qualitative result.} Even though the model in (\ref{liuwa}) can be derived from a weak coupling analysis of the gauge theory, it can be thought of as a phenomenological model describing supergravity in $AdS_{5}$.
\par
 We are motivated to discuss the fixed charge ensemble in the same spirit. Let us add a small interaction term to (\ref{seff}). The effective action is then given by
\bea
S_{eff}(\rho)=N^{2}(a\rho^{2}+b\rho^{4}+q(\sqrt{(1+\frac{c^{2}}{q^{2}}\rho^{4})}+\log(\frac{\frac{c}{q}\rho^2}{1+\sqrt{1+\frac{c^{2}}{q^{2}}\rho^{4}}})))+O(1)
\label{SeffInt}
\eea
Here $b$ is proportional to $\lambda$ and is also a function of charge. Depending on the theory considered, the sign of $b$ can be either positive or negative. It has been shown in \cite{Aharony:2005bq} that $b$ is positive for a pure YM theory. In the following discussion we will assume that this is the case in order to motivate a similarity with the supergravity picture.
The equations determining the saddle points of (\ref{SeffInt}) , including the contribution from the path integral measure, are 

\bea
\nn (a+c)\rho^2+2b\rho^4+\frac{q}{(1+\frac{c^2}{q^2}\rho^{4})^{\frac{1}{2}}+\frac{c}{q}\rho^{2}}  & = &  \rho^2, \rho <\frac{1}{2} \\
    (a+c)\rho^{2}+2b\rho^{4}+\frac{q}{(1+\frac{c^2}{q^2}\rho^{4})^{\frac{1}{2}}+\frac{c}{q}\rho^{2}} & = & \frac{\rho}{4(1-\rho)}, \rho > \frac{1}{2}
\label{TosolveInt}
\eea
\par
In what follows it will be useful to introduce a function  
\bea
\nn M(\rho) & = & \rho^{2}, \rho <\frac{1}{2} \\
        & = & \frac{\rho}{4(1-\rho)},\rho>\frac{1}{2}    
\eea
$M(\rho)$ is an increasing convex function and is the right hand side of (\ref{TosolveInt}). It is also useful to introduce $D(x)=S'_{eff}(x)-M(x)$. Eqn (\ref{TosolveInt}) is then equivalent to $D(\rho)=0$. 
\par
It has been shown in \cite{Alvarez-Gaume:2005fv} that for an interacting model with zero charge, we get nucleation of black holes along the curve given by $a=\frac{1-2w}{(1-w)^{2}(1+w)}$ and $b=\frac{2w}{(1-w)^{2}(1+w)^{3}}$, $1>w>0$. Here we want to analyze a similar type of phenomenon. Let us consider the different cases.
\par

\subsection*{\underline{$T$ is varying and $q$ is small}}
\par 

Let us start with a value of charge which is small (i.e. $q \ll 1$) and let us increase the temperature from zero. At low temperatures all the parameters $a, c$ will be small (for small $T$ these parameters have a dependence like $e^{-c\beta}$, $c$ is a constant ). Hence we get just one solution for $\rho < \frac{1}{2}$ which we call $I$. There is no solution for $\rho > \frac{1}{2}$ (see the topmost curve of Fig \ref{Fixedq}) because the left hand side of (\ref{TosolveInt}) is less than the right hand side.
\par 
The situation is quite similar to supergravity where for small charge and low enough temperatures we get a stable small black hole solution. \footnote{We should keep in mind that $\rho < \frac{1}{2}$ is not the supergravity regime. The solutions of gauge theory effective action there should be represented as excited string states.}

\begin{figure}[h!]
\begin{center}
\includegraphics[height=2in]{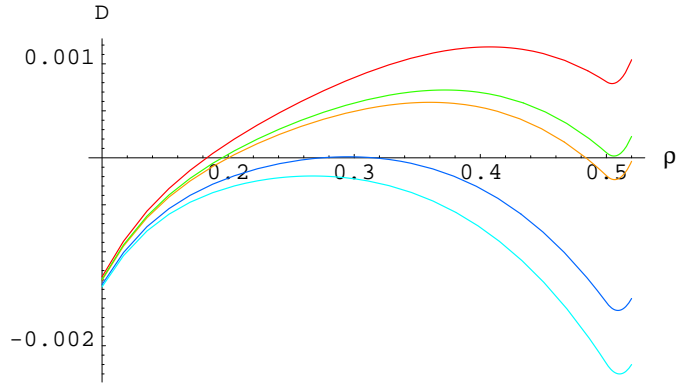}
\end{center}
\caption{Plots of $D(\rho)$ with increasing $a=c$ from the above and with fixed $b$ and $q<q_{crit}$}
\label{Fixedq}  
\end{figure}

The function $M(\rho$) (i.e. right hand side of (\ref{TosolveInt})) is a convex increasing function. Hence we will generate new solutions of (\ref{TosolveInt}) in the region $\rho > \frac{1}{2}$ as we increase temperature (i.e. $a(T),c(T)$ as discussed in appendix \ref{app:genprop}) keeping $q$ fixed (\fig{Fixedq}, \fig{Fixedq1}). The new solutions will always come in pairs (Fig \ref{Fixedq}). Let us call the solution nucleation temperature as $T_{01}(q)$. At $T=T_{01}$ we have
\bea
\nn D(\rho)=S_{eff}'(\rho)-M(\rho) &=& 0 \\
    D'(\rho)=0
\eea

\begin{figure}[h!]
\begin{center}
\includegraphics[height=2in]{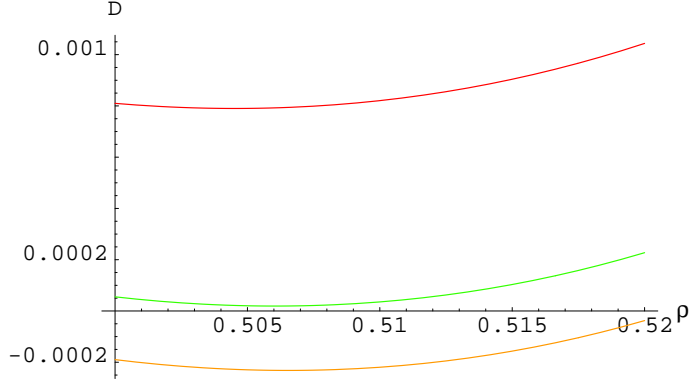}
\end{center}
\caption{ Top three graphs of Fig $2$ showing emergence of two saddle points in the region $\rho > \frac{1}{2}$ } 
\label{Fixedq1}    
\end{figure}
\par
As we increase the temperature further (i.e. $a(T), c(T)$) the function on the left hand side of (\ref{TosolveInt}) will also increase. Hence the solutions will start to separate. Let us call them $II$ and $III$. Here $\rho_{III}>\rho_{II}$. Also, $III$ is a stable saddle point whereas $II$ is an unstable one. They are similar to stable big and unstable small black holes in supergravity. As the temperature is increased beyond $T_{01}$ the value $\rho_{II}$ decreases whereas $\rho_{III}$ increases. At some temperature $T_{H}(q)$, we will have $S(III)<S(I)$ and consequently we expect a first order phase transition.  At a temperature $T_{HP}$, the dominant saddle point of the system changes from $I$ to $III$. As the temperature increases the saddle point $II$ goes through a third order transition when $\rho(II)$ crosses the $\rho={1 \ov 2}$ point. Call this temperature $T_{3}$ which is determined by the following relation between the parameters

\bea
(a+\frac{b}{2})+4q(1+\frac{c^{2}}{16q^{2}})^{\frac{1}{2}}=1
\eea
\par
Increasing the temperature further, the saddle point $II$ approaches the saddle point $I$ and they merge  at a temperature $T_{02}$ ($4th$ graph from above in Fig \ref{Fixedq} ). In the language of catastrophe theory this is a fold
 catastrophe. For $T>T_{02}$, the only saddle point is $III$. This then is the thermal history as we increase the temperature for a small $q$.
\par
In summary at low temperatures, we have only one saddle point $I$ and then two new saddle point $II,III$ are created at $T_{01}$. As the temperature increases the saddle points $I , II$ merge at a temperature $T_{02}$. Beyond that we have only one saddle point $III$. In the next paragraph we will discuss what happens when we increase the value of the $q$.


\subsection*{\underline{Varying q$:$}}
\par
     Let us discuss how the various temperatures discussed above change as we increase the value of $q$. 
\begin{enumerate}

\item The first is $T_{01}(q)$, the nucleation temperature for saddle points $II$ and $III$. $T_{01}$ will decrease as we increase the value of $q$. This is so because all three terms in the right hand side of (\ref{TosolveInt}) are positive and increasing functions of $\rho$ and the left hand side is a positive convex function. Consequently the saddle point value of $\rho$ at $T_{01}$ will also decrease.

\item The temperature $T_{02}$ at which the stable saddle point $I$ and unstable saddle point $II$ merge and the value of $\rho$ at $T_{02}$, will increase as we increase $q$. The reason is that all the coefficients in (\ref{TosolveInt}) are positive and hence increasing the value of $q$ increases the function on the left hand side.

\end{enumerate}

\par
 As we increase the value of $q$ further, $T_{01}$ and $T_{02}$ will become equal for some value of $q=q_{crit}$. In the language of catastrophe theory this is a cusp catastrophe. Corresponding to $q_{crit}$ there will be a $T_{crit}=T_{01}=T_{02}$ and a value of $\rho_{crit}$ (saddle point $\rho$ at $q_{crit}$ , $T_{crit}$).  As we increase the $q$ beyond $q_{crit}$, we do not get any new saddle point and consequently  there is only one saddle point for all temperatures. We will discuss the physics near this phase transition in detail in what follows. 
\par

\subsection*{\underline{Value of $\rho_{crit}$ :}}

Let us determine the value of $\rho_{crit}$. As already discussed at the end of the previous chapter, in (\ref{expansion}) $C$ is always a finite quantity. Hence, (\ref{TosolveInt}) cannot have three solutions in the region $\rho<\frac{1}{2}$ for small $b$. Therefore the saddle point $I$ is always in the region $\rho < \frac{1}{2}$. Whereas the saddle point $III$ will be in the region $\rho>\frac{1}{2}$. Hence the only place where these three saddle points can meet is $\rho_{crit}=\frac{1}{2}$ which is also the point of the third order phase transition.

\vspace{0.5cm}
\subsection*{\underline{Physical Interpretation}}
\par
Before proceeding further we will briefly discuss the bulk interpretation of saddle points of weakly coupled gauge theory. Weak coupling in gauge theory means $l_{s}\gg R_{AdS_{5}}$. Hence the supergravity picture is not valid in the bulk. However at large $N$, the string coupling (i.e. $\frac{1}{N}$) will be small and we may conclude that the saddle points discussed above can be described by exact (in all orders in $l_{s}$) conformal field theories. These CFTs  are characterized by the values of  $q$ and $\rho$ at the saddle points.

\par 

We would like to end this section by emphasizing that the coincidence of the three saddle points at $\rho=\frac{1}{2}$, is a property of the weak coupling ($\lambda << 1$) limit of the gauge theory. In what follows we will see that this fact is not necessarily true at strong coupling. We will show that there the coincidence happens in the gaped phase where $\rho>\frac{1}{2}$.

\section{Effective action and phase diagram at strong coupling} \label{sec:simmod}

In this section we will discuss the effective action and the phase diagram in the strongly coupled gauge theory which is dual to the supergravity (discussed in section \ref{Supergravity}) regime of $IIB$ string theory. 
\subsection{ Finite temperature effective actions in the gauge theory}
\par
Let us first summarize the situation in the zero charge sector. The propagator of adjoint fields in the free gauge theory, on a compact manifold $S^{3}$, coupled to a space independent $A_{0}$ is given by (see \cite{Kazakov:2000pm})   
\bea
G^{ij}_{U ~ kl}(x,t,y,0)=\sum_{n=-\infty}^{n=\infty} (U^{n})^{i}_{j} G_{0}(x,t+n\beta,y,0) (U^{-n})^{k}_{j} 
\label{green}
\eea 
 where $G_{0}(x,t,y,0)$ is the zero temperature Green's function and $U$ is the constant Polykov line. We know that at any temperature \footnote{Temperature here is measured in units of $\frac{1}{R_{S^{3}}}$.} $G_{0}(x,t+n\beta,y,0) > G_{0}(x,t,y,0) $ and also at low temperatures
\bea
G_{0}(y,t+n\beta,x,0) \sim e^{-n\beta} 
\label{lowtemp}
\eea
Using the above Green's function one can develop the large N diagrammatic to arrive at an effective action involving $Z_N$ invariant terms built out of products of $trU^n$. In fact one can imagine integrating out all the modes $trU^n$ for $n>1$ in favor of $trUtrU^{-1}$. In this way one gets an effective action of the form 
\bea
S_{eff}=\sum_{n=1}^{n=\infty} a_n(\beta,\lambda) (\frac{TrUTrU}{N^2})^n
\eea

As one increases the coupling constant we would expect that the form of the effective action would remain the same except that the dependence of the parameters on temperature and the 'thooft coupling would change. 
\subsection{Non-zero charge sector}
\par
If we include the fixed charge constraint, as in (\ref{chargefix1}), then we get the following expression for the fixed charge path integral  
\bea
Z_{q}=\int d\mu e^{i\mu Q}\int DU e^{N^{2}(\sum_{n=0}^{n=\infty} S_{n}(\rho,\lambda,\beta) \cos(n\mu))}   
\label{stract}
\eea
For large $N$ we can do the $\mu$ integral by the saddle point method. The saddle point of $\mu$ is on the imaginary axis. Hence we set $\mu=im$, to get the saddle point equation 
\bea
q=\sum_{n=1}^{n=\infty} nS_{n}(\rho,\lambda,\beta) \sinh (nm)), q=Q/N^{2}
\label{GenSaddle}
\eea 
At small values of $\rho$, $S_{n}(\rho)$ goes as $\rho^{n}$ \footnote{Introduction of a chemical potential changes the formula (\ref{green}) as 
\be
\nn G^{\mu ~ ij}_{U ~ kl}(x,t,y,0)=e^{\frac{i\mu t}{\beta}} \sum_{n=-\infty}^{n=\infty} (U^{n})^{i}_{j} G^{\mu}_{0}(x,t+n\beta,y,0) (U^{-n})^{k}_{j}
\ee
 where
\be
G^{\mu}_{0}(x,t+n\beta,y,0)=e^{i n \mu}G_{0}(x,t+n\beta,y,0)
\ee
Hence in each order in perturbation theory the terms containing $\cos(n \mu)$ also get multiplied by 
$(TrUTrU^{-1})^{m},m>n$ or the higher operators like $TrU^{n}$ which can be integrated out to give again a term like $(TrUTrU^{-1})^{m}$. 
  } and hence in the $\rho \rightarrow 0$  limit  we can approximate the equation (\ref{GenSaddle}) as 
\bea
q \approx \CC \rho \sinh (m)
\eea
where $\CC$ is a constant independent of $\rho$.
The solution is
\bea
m \approx q\log(\rho)+\CC , \rho \rightarrow 0
\eea
Substituting $m$ in ($\ref{stract}$) we get a logarithmic term for $\rho$. We conclude that the logarithmic term is a general feature and it implies among other things that $TrU=0$ is never a solution in the non-zero charge sector.

\subsection{Model effective action at strong coupling}
\par   
Following our previous discussion, we will include the generic logarithmic term in the effective potential for a fixed non-zero charge, and proceed to analyze the saddle point structure following \cite{Alvarez-Gaume:2005fv}.

Our proposal for the gauge theory effective action is 
\bea
S_{q}=S(a(T),b(T),...,\rho)+q\log(\rho)
\eea
and the saddle point equations are     
\bea
\nn \rho F(a(T),b(T),c(T),...\rho^2)+q  =\rho^{2} , \rho <\frac{1}{2} \\
\rho F(a(T),b(T),c(T),...\rho^2)+q =\frac{\rho}{4(1-\rho)},\rho>\frac{1}{2}
\label{TosolveIntStr}
\eea
Where $F(\rho)=S'_{eff}(\rho)$. We assume that 
\begin{enumerate}
\item{ $F(x,T)$ is a monotonically increasing function of $x$ and F(0,T)=0}
\item{ Value of $F(x,T)$ increases for fixed $x$ as we increase the temperature and F(x,0)=0.}

These global properties of $F(x)$ reproduce the phase diagram of supergravity.

\end{enumerate}   
\par
\subsection*{\underline{Analysis of solution structure}}

\par
     Let us consider the function $D(T,\rho)=\rho F(\rho,T)-M(\rho)$ (Where $M(\rho)$ is the contribution from measure appearing at the right hand side of (\ref{TosolveIntStr})). At $T=0$, $F(\rho,T)$ is zero. Hence  $D(T,\rho)$ is a monotonically decreasing function of $\rho$ at $T=0$.
\par
 We know that at $T=T_{01}$ a pair of  two new saddle points appear at $\rho_{01}>\frac{1}{2}$. Hence at $T=T_{01}$ we have $D(T_{01},\rho_{01})=0$ and $D'(T_{01},\rho_{01})=0$. $D(T_{01},\rho)$ has a zero for $\rho=0$   and it is a decreasing function in the neighborhood of $\rho=0$. It again increases and become zero at $\rho_{01}$ and then the function again decreases as $\rho \rightarrow 1$. This implies that the function has a local maximum and local minimum. 

\begin{figure}[h!]
\begin{center}
\includegraphics[height=2in]{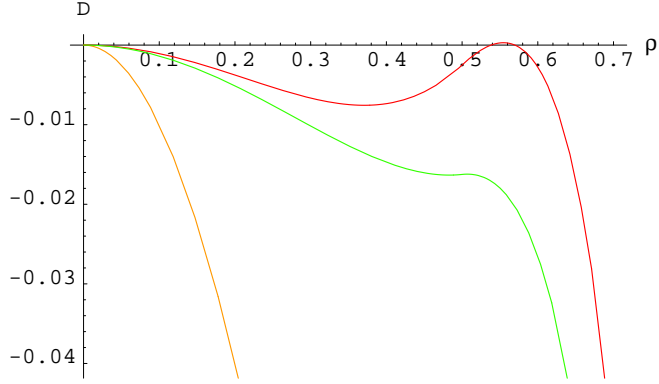}
\end{center}
\label{Emergence}    
\caption{Plots of $D(T,\rho)$ for $T=0$, $T=T_{crit}$,$T=T_{01}$ from below.} 
\end{figure}

\par
In summary:
\begin{enumerate}
\item{$D(0,\rho)$ is a monotonically decreasing function of $\rho$.}
\item{$D(T_{01},\rho)$ has a maximum and minimum.}
\end{enumerate}
\par
There is a temperature $T_{crit}$ at which the local maximum and local minimum appear (Fig \ref{Emergence}). Let us call this temperature $T_{crit}$. At $T_{crit}$ the curve $D(T_{crit},\rho)$ will have a point of inflection at $\rho=\rho_{crit}$, say. Let the value of $D(T_{crit},\rho_{crit})=q_{crit}$.
\par
Increasing the  value of $q$ from zero we need to solve the equation $D(T,\rho)=q$. We will get a solution for a non-zero value of $\rho$. Denote this solution as $I$. As the temperature increases, two new solutions  appear at $T=T_{01}$. Call the stable solution as $III$, and the unstable solution $II$. As  the temperature is further increased to $T_{02}$, the unstable solution $II$ and the stable solution  $I$ merge.  For $T>T_{02}$ , the only solution is $III$. 
 \par
 As $q$ approaches $q_{crit}$ from below the two temperatures $T_{01}$ and $T_{02}$ approach each other.  At $q=q_{crit}$, we have $T_{01}=T_{02}=T_{crit}$. If we increase $q$ beyond $q_{crit}$ only one solution appears for all temperature. These facts are consistent with supergravity solutions(section \ref{Supergravity}).
\par
 With a sufficiently sharp rising function $F(T,\rho)$ in (\ref{TosolveIntStr}) we can obtain this critical point in the region $\rho>\frac{1}{2}$.\footnote{Which describes supergravity in the bulk\cite{Alvarez-Gaume:2005fv}.}. As the function $D(\rho,T)$ is smooth in the region $\rho>\frac{1}{2}$ the second derivative of the function $D(T,\rho)$ will vanish at the inflection point, and we will get a third order phase transition. We can calculate the partition function in suitable double scaling limit near the critical point. This is discussed in section \ref{Unin}.
\par
\subsection*{\underline{A specific example}}

We will now illustrate the above phenomenon in a simple model defined by 
\bea
F(\rho)=a\rho+b\rho^{3}
\label{newmod}
\eea
where $a,b > 0$.

We will determine the parameter ranges of $a,b$ for which all the three saddle points of (\ref{newmod}) are in the range $\rho>\frac{1}{2}$.

\begin{figure}[h!]
\begin{center}
\includegraphics[height=2in]{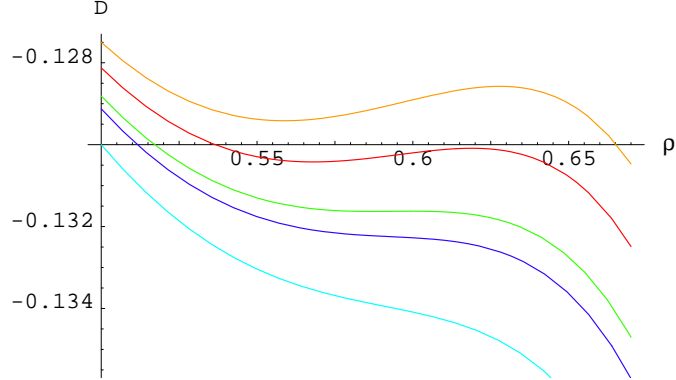}
\end{center}
\label{Tricritical}    
\caption{Plots of $D(T,\rho)$ with with fixed $a$ and increasing $b$ from the top, showing a critical transition in the region $\rho>\frac{1}{2}$ (from the top, the $3rd$ graph has a point of inflection) }
\end{figure}

At $\rho=\frac{1}{2}$ we have the constraints  

\bea
\nn \del_{\rho}(\rho F(\rho)-\frac{\rho}{4(1-\rho)})<0 \\
\del_{\rho}^{2}(\rho F(\rho)-\frac{\rho}{4(1-\rho)}>0 
\label{rho}
\eea
Putting the value of $\rho=\frac{1}{2}$ in the above inequality we get the following constraints on the parameters $a+b<1$ and $a+3b>2$. Simplifying we have $b>\frac{1}{2}$ and $a<\frac{1}{2}$.

 \par
As the coupling become stronger, we expect that $b$ is not necessarily small and will be of $o(1)$ or greater. All the saddle points of (\ref{TosolveIntStr}) are then naturally shifted to the region $\rho>\frac{1}{2}$. Here, as was discussed in \cite{Alvarez-Gaume:2005fv}, we can expect to match the solutions of the gauge theory with those of supergravity. The stable saddle point $I$ corresponds to the stable black hole branch $I$ of supergravity. And unstable saddle point $II$ is matched with the unstable black hole branch $II$. The stable saddle point $III$ is matched with the big stable black hole in supergravity. With this identification the thermal history and critical behavior of the gauge theory, discussed earlier in this chapter, match with the thermal history and critical behavior of supergravity (discussed in section \ref{Supergravity} and \cite{Chamblin:1999hg}).

\section{Universal neighborhood of critical point and the critical exponents}\label{Unin}

Let us consider the effective action $S_{tot}(\rho,T,q)$ which includes the contribution from the path integral measure over an unitary matrix. The derivative of $S_{tot}$ with respect to $\rho$, say $G(\rho,T,q)$, gives the saddle point equations (\ref{TosolveIntStr}). We have already discussed that by a suitable choice of parameters the critical point appears in the region $\rho>\frac{1}{2}$. This critical point is a third order critical point because three saddle points of the system merges here. Hence the first and second derivatives of $G_{tot}(\rho,T,q)$ with respect to $\rho$ vanish at $\rho=\rho_{crit}, q=q_{crit},T=T_{crit}$. Expanding $G(\rho,T,q)$ around the critical point , we get
\be
G(\rho,T,q)=(\delta \rho)^{3}\frac{\del^{3}_{\rho}G}{3!}+(\delta T)\del_{T}G+(\delta q)\del_{q}G+(\delta\rho)(\delta T)\del_{\rho}\del_{T}G+(\delta q)(\delta T) \del_{\rho}\del_{q}G
\label{Geq}
\ee

 \par
Let us fix $T=T_{crit}$ or $\delta T=0$. Then the equation (\ref{Geq}) has one solution. In order to know how the saddle point value of $\rho$ approaches $\rho_{crit}$ ($\delta \rho \rightarrow 0$) as $\delta q \rightarrow 0$, we equate the leading part of (\ref{Geq}) to zero.
\bea
(\delta \rho)^{3}\frac{\del^{3}_{\rho}G}{3!}+(\delta q)\del_{q}G=0
\eea
Hence $\delta \rho \propto \delta q^{\frac{1}{3}}$ and we get the same universal exponent $\frac{1}{3}$ as in supergravity \cite{Chamblin:1999hg}. 

\subsection{Partition function near the critical point}
Near the critical point we can write the $S_{tot}$ as    
\bea
S_{tot}=S_{tot}(\rho_{crit},T_{crit},q_{crit})+(\delta\rho)^{4}\frac{\del^{4}_{\rho}S}{4!}+(\delta q )\del_{q}S+(\delta q)(\delta \rho)\del_{\rho}\del_{q}S+O(\delta \rho^5)
\label{Seq}
\eea
If we define a double scaling limit $N^\frac{1}{2}\rho=x,N^{\frac{3}{2}}q=z$ , we can write the $o(1)$ part of the partition function, after suitable rescaling of the variables, as 
\bea
Z_{2} \propto && \int dx e^{-(x^4-zx)}
\eea 
This can be calculated in a power series 
\bea
Z_{2} \propto \sum_{n=0}^{\infty} \frac{z^{2n}}{(2n)!} \Gamma (\frac{n}{2}+\frac{1}{4})
\eea 

\subsection{Approaching the critical point through a line of first order transitions}
\par
Another type of double scaling limit is possible in this problem. We can set\footnote{It is same as following the HP(first order) transition line in parameter space.} 
\bea
(\delta T)\del_{T}G+(\delta q)\del_{q}G=0 
\label{rel}
\eea
by choosing a suitable relation between $\delta T$ and $\delta q$. Using (\ref{rel}) in (\ref{Geq}) we get  
\bea
 (\delta \rho)^{3}\frac{\del^{3}_{\rho}G}{3!}+(\delta\rho)((\delta T)\del_{\rho}\del_{T}G+(\delta q)\del_{\rho}\del_{q}G)=0
\eea

with the solutions 
\bea
\nn \delta \rho=0 \\
\delta \rho \propto \pm(\delta T)^{\frac{1}{2}}
\label{prop}
\eea

We can expand $S_{eff}$ as 
\bea
S_{eff} \approx S_{crit}+(\delta\rho)^{4}\frac{\del^{4}_{\rho}S}{4!}+C_{1}(\delta T)(\delta \rho)^{2}+ OT
\eea 
where $OT$ are terms independent of $\delta\rho$. Defining a suitable double scaling limit. $N^{\frac{3}{2}}\delta T=z, N^{\frac{1}{2}}\delta\rho=x$ and a suitable rescaling of the parameters we can evaluate the $o(1)$ factors in the partition function as 
\bea
Z_{2} \propto \int dx e^{-(x^{4}+2 z x^2)} \propto \sum_{n=0}^{\infty} \frac{(2z)^{n}}{n!} \Gamma (\frac{n}{2}+\frac{1}{4}) 
\propto \sqrt{z}e^{\frac{z^{2}}{2}}K_{\frac{1}{4}}(\frac{z^{2}}{2})
\eea

where $z<0$.



\section{Conclusion}
\par  
In this paper we have studied the logarithmic matrix model generated by fixing the R-charge in the gauge theory partition function. In the free gauge theory it has been shown that there is no solution with $\rho=0$ ($AdS$ type solution). We then studied the effect of adding an interaction term in our model and discussed the generic nature of the logarithmic term even at arbitrary value of the coupling. We identified the supergravity saddle points and their critical behavior which was discussed in (\cite{Chamblin:1999hg}). 
\par
  Our main aim was to give another example of the utility of unitary matrix methods in providing a non-perturbative dual description of blakholes in $AdS$ and to understand the relation between matrix models and string theory in general. It would be interesting to consider an effective unitary matrix model to describe phases of Kerr-Ads black holes.    

\section{Acknowledgment}
 We would like to thank the theory division of CERN for hospitality where part of this work was done. We acknowledge useful discussions with Luis Alvarez-Gaume and Marcos Marino on many aspects of the matrix model/string theory correspondence. We also acknowledge a correspondence with Hong Liu. PB likes to acknowledge CSIR for SPM fellowship and TIFR alumni association for partial financial support. PB also likes to thank Swagato Mukherjee for technical help in preparation of this paper.

\appendix

\section{Appendix: Inclusion of Fermions}\label{app:fermion}
Including the contributions from the fermions of $N=4$ $SYM$ theory change (\ref{IntegralZ}) to 
\bea
 Z(\beta,Q_{0}) =  \int DU \int d\mu  \exp(N^{2}(a+ c \cos(\mu)+d \cos({\mu \ov 2}))\rho^{2} -i\mu Q_{0}) 
\label{IntegralF}
\eea 
Where $d(\beta)$ is the single particle partition function for the fermions.

At large $N$ , the integral in (\ref{IntegralF}) could be evaluated by the saddle point method. The equations determining the saddle points of $\mu=im$ and $\rho$ are

\be
(c \ \sinh(m)+{d \ov 2} \sinh({m \ov 2})) = {q \ov \rho^{2}} 
\label{SadM}
\ee
and 
\be
\rho (a+\cosh(m)+d\cosh({ m \ov 2})) = \rho 
\label{SadR}
\ee 
We would like to see weather there is a solution with $\rho=0$. As the
right hand side of (\ref{SadM}) becomes large in the limit $\rho \rightarrow 0$, we can self consistently approximate $\cosh(m)$ and $\sinh(m)$ as $e^{m}$ and we get 
\be
m \approx \log{q \ov c \rho^{2}} 
\label{sol}
\ee
Hence a logarithmic potential for $\rho$ is once again generated. One can also confirm this by putting (\ref{sol}) in (\ref{SadR}).

\section{Appendix: Positivity of the coefficient of the quadratic term in the effective action}\label{app:genprop}

Let us consider the partition function of YM theory on a compact manifold written as an integral over the effective action of $\rho=TrUTrU^{-1}$.
\bea
 Z(\beta)=\int DU e^{N^{2}(S_{eff}(\rho))} \\
=\int d\rho e^{N^{2}(S_{eff}(\rho)-S_{M}(\rho))}
\label{part}
\eea
Where $S_{M}(\rho)$ is the contribution from the measure part\footnote{See discussions before (\ref{Tosolve1}).} of path integral and
\bea
S_{eff}(\rho)= a({\beta})\rho^{2}+\sum_{n=4}a_{n}(\beta)\rho^{n}
\eea
 i.e. a polynomial in $\rho$.
As $\beta \rightarrow \infty$ we have $S_{eff}(\rho) \rightarrow 0$. Contribution from the measure part $S_{M}(0)$ has only one minimum at $\rho=0$. Hence at low temperature the system will have a saddle point at $\rho=0$. Expanding $\rho$ around this saddle point as $\rho=0+\frac{\delta\rho}{N}$ we get 
\bea
Z(\beta)=\int_{-\infty}^{\infty} d(\delta \rho) e^{-(1-a(\beta))(\delta \rho)^{2}+o({1 \ov N^2})}
\label{part2}
\eea 
Like any thermal partition function, (\ref{part}) or (\ref{part2}) is a decreasing function of the $\beta$. Hence $a(\beta)$ should also be a decreasing function of $\beta$. Since $a(\infty)=0$, for any finite $\beta$, $a(\beta)$ is a positive decreasing function. Hagedorn transition happens when $a(\beta_{H})=1$, but whether $a(\beta)$ will reach $1$ or not depends on the model.

\end{document}